\documentclass[preprint,11pt]{elsarticle}

\usepackage{hyperref}
\usepackage{balance}
\usepackage{multicol}
\usepackage{multirow}
\usepackage{tabularx}
\usepackage{amsmath}
\usepackage{float}
\usepackage{lineno}
\usepackage{soul}
\usepackage{makecell}
\usepackage{color}
\usepackage{booktabs}

\graphicspath{{figures/}}

\begin{document}

\begin{frontmatter}
\title{A fully autonomous data center for the space-borne hard X-ray Compton polarimeter POLAR developed at PSI}
\author[psi]{Hualin Xiao \fnref{fhnw}}
\author[psi]{Wojtek Hajdas \corref{cor}}
\ead{wojtek.hajdas@psi.ch}
\author[psi]{Radoslaw Marcinkowski\fnref{radecs}}
\cortext[cor]{Corresponding author}
\fntext[fhnw]{Present address: University of Applied Sciences and Arts Northwestern Switzerland (FHNW), Windisch, Switzerland}
\fntext[radecs]{Present address: RADEC, Koblenz 5322, Switzerland}
\address[psi]{Paul Scherrer Institute, 5232 Villigen PSI, Switzerland}

\begin{abstract}
Constant data inflow from the hard X-ray polarimeter POLAR onboard of the Chinese Space Laboratory TG-2 requires fully automated and safe guarded data processing.
For this purpose, a dedicated data centre was established at PSI.
This paper presents its design concept and structure as well as demonstrates its main features
with respect to data products, data processing pipelines, quick-look utilities and web-based applications.
As POLAR telemetry data reaches as large as 100 GB daily, specialized software and hardware with dedicated architecture are required to store and process this huge amount of data.
POLAR is a space-borne detector dedicated for precise measurements of the linear polarization of hard X-rays emitted from solar flares and gamma-ray bursts in the energy range of 50 keV to 500 keV.
It was launched into a low earth orbit on September 15th, 2016.

\end{abstract}

\end{frontmatter}

\section{Introduction}

POLAR is a space-borne hard X-ray Compton polarimeter built through a collaboration
of institutes from Switzerland, China and Poland. It is equipped with an array of 1600 scintillator bars.
Such a detector configuration allows for precise measurements of the polarization of hard X-rays.
POLAR was optimized for the  50 keV to 500 keV energy range.
It has a large field of view of about 1/3 of the sky and a sensitive area of about 80 cm$^2$.
Its minimal detectable polarization (MDP) for stronger Gamma-Ray Burst (GRB) events is lower than 10\%.
Polarization detection of solar flares can be determined with even better accuracy.
POLAR was launched into space on an orbit around 380 km on September 15th, 
2016 on-board the Chinese Space Laboratory TG-2 for up to a 3-year long observation period.

During its normal space operation, POLAR acquires data continuously apart from short 
passages through the South Atlantic Anomaly (SAA).
POLAR detects about ten GRBs and slightly more of the solar flares per month. 
Therefore the telemetry is dominated by background events accounting for up to tens of thousands events per second.
As a result, the daily telemetry datasets are as large as about 100 GB threreabouts.
Both dedicated hardware and specialized software are required to receive,
analyze and archive such a huge amount of data. A constant data flow also requires fully automated and safeguarded data processing.
To realize these goals, a dedicated data center was established at the Paul Scherrer Institute (PSI).
A suite of software applications as well as an instrument database were developed for data processing, quick-look analysis and alerting purposes.
In this paper we present both the design concept and the structure of the POLAR data center at PSI (PPDC) including detailed descriptions of its main features.

\section{POLAR instrument}
\subsection{POLAR instrument}

\begin{figure*}[htb]
\begin{minipage}{0.45\linewidth}
\includegraphics[width=0.95\textwidth]{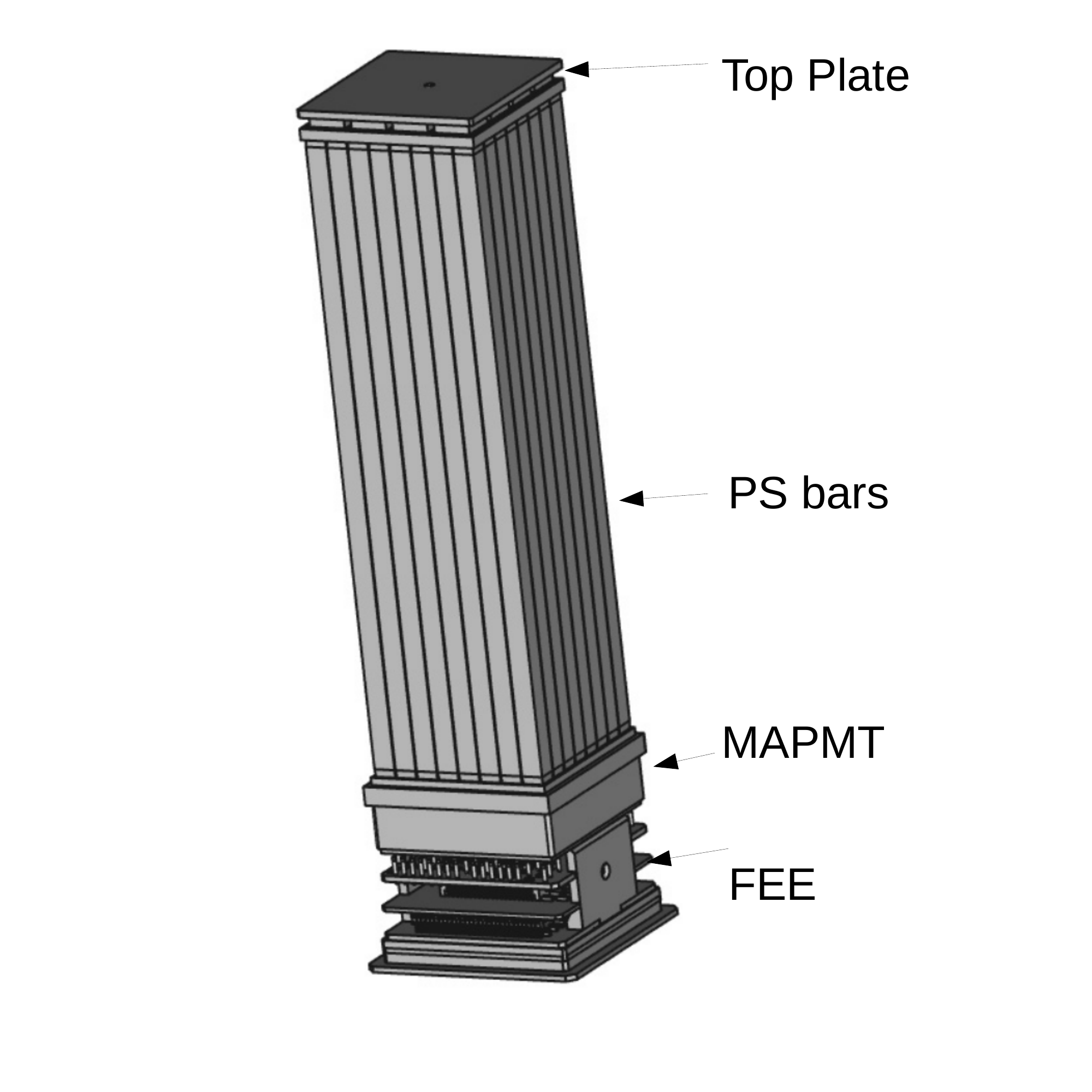}
\end{minipage}
\begin{minipage}{0.5\linewidth}
\includegraphics[width=0.95\textwidth]{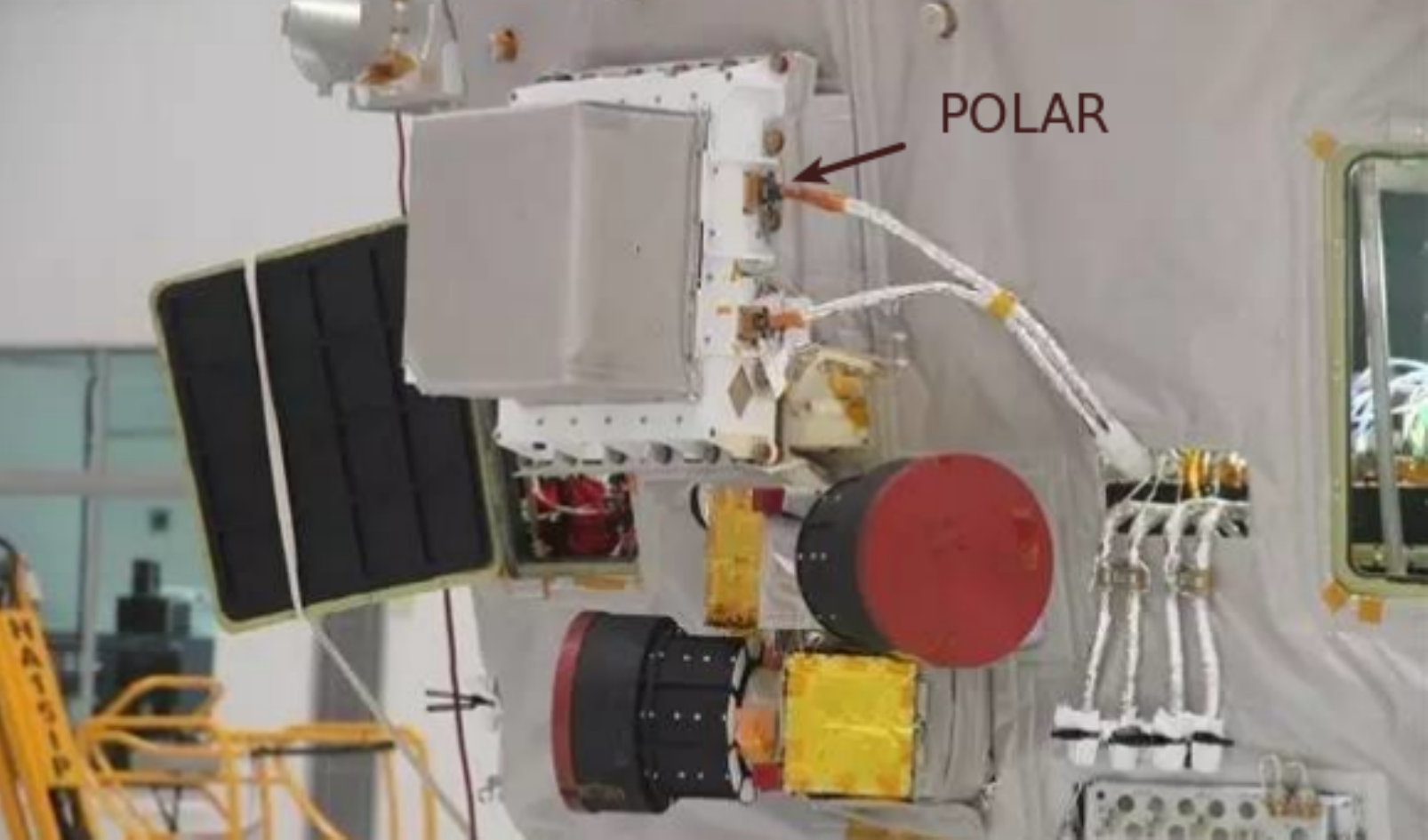}
\end{minipage}
\caption{POLAR detector module structure (left) and a photo of the POLAR flight model on TG-2 (right).
(Note that the carbon fiber socket of the module is not shown in the left panel.)
}
\label{fig:polar}
\end{figure*}

The POLAR instrument consists of 25 identical modules placed inside the carbon fiber enclosure.
The modules contain in total 1600 plastic scintillator (PS) bars.
Each module consists of 8$\times$8 PS bars, a 64-pixel multi-anode photo-multiplier (PMT) 
(Hamamatsu MAPMT H8500) and a readout front-end electronics (FEE) as shown in the left panel of Fig.~\ref{fig:polar}.
Each bar has  dimensions of $5.8\times5.8\times176 $ mm$^3$. 
The 64 bars are fixed together with two 8 $\times$ 8 plastic grids and
coupled to the MAPMT.
The FEE processes signals from the MAPMT, communicates with the POLAR Central Task Processing
Unit (CT) and provides distributed high voltage to the MAPMT dynodes.
The fast processing logic of the FEE packs together signals from event digitized
energy depositions and triggers hit patterns and includes timestamps as well as some 
auxiliary data (such as the FEE health status). The above data forms a module science packet that is sent to the POLAR CT.
The CT controls and monitors all 25 detection modules and is responsible for trigger decisions based on pre-analysis of incoming events. It also handles communications with the POLAR IBOX (acting as an interface to the TG2 space-lab) and manages both the low voltage power supply (LVPS) and
the high voltage power supply (HVPS).
All 25 modules, the CT, the power supplies and the interfacing electronics are enclosed in a
carbon box (OBOX) shown on the right panel of Fig.~\ref{fig:polar}.
The whole instrument is mounted on the outside panel of the space-lab and faces the sky permanently.
Calibration of POLAR is realized using four weak $^{22}$Na  sources installed at  the innermost edges of its four corner  modules. 
The $^{22}$Na nuclei emit two collinear photons with energies of 511 keV.
Selection of these two collinear photons during off-line analysis allows for proper calibration of the  whole  detector \cite{hlastro2018}. 

POLAR has six operation modes: 1) initialization, 2) standby, 3) maintenance , 4) normal, 5) SAA and 6) diagnostic. 
After being powered on, its registers in the CT and the FEEs are configured as programmed in
the initialization procedure.
If no telecommand occurs, POLAR switches 
to the normal mode shortly after initialization.
In this mode, the instrument  collects data 
continuously apart from short passages through the South 
Atlantic Anomaly (SAA) where the normal mode is temporarily replaced by the SAA mode.
More details on the POLAR
instrument and its operations can be found in Refs., e.g., \cite{hlastro2018,nicolas, silvio, hlastro2016}.

\section{Hardware at the PSI POLAR data center}

The PPDC relies on the high-performance computing resources installed on the PSI IT premises.
These resources consist of two powerful servers dedicated to data processing and data archiving.
The processing server fulfills  heavy duty and intense computing requirements; it is  equipped
with 16 CPU cores and 64 GB of RAM, 2 TB SSD and a 16 TB hard disk.
The archiving server currently uses an expandable disk array with a total space of 64 TB.
These two servers run a Linux operating system and are connected together via the Gigabit Ethernet.
The hard disks in both servers use a Network File System (NFS) and have been configured as a 
Redundant Array of Independent Disks Level 5 (RAID5).
The RAID5 configuration provides sufficient protection for the data
even in the case of a failure of the whole physical drive.
Another smaller Linux server is  installed as a backup for the processing and storage units. 
It is also used for tests during development of new and updated software solutions.

\section{POLAR raw data and data transmission}

The main raw data received at the PSI POLAR data center (PPDC) falls into
three categories: science data, housekeeping data (also called engineering data)
and platform data (see Table \ref{tb:types}).

There are two different types of science data: module science data and trigger data.
The module science data is generated by detector modules. 
It contains event trigger timestamps, energy depositions in its channels 
(i.e., ADC channels) and 64 trigger pattern bits.
The trigger data generated by the CT, contains the collected trigger patterns from 25
modules, dead time counter, trigger counter and all timestamps.
Such a configuration is necessary to attribute and merge the science packets from different modules to the same detected events.
The  size of the  daily science data received at the PPDC ranges from  10 GB to 100 GB, with an average of 50 GB.
It depends not only on the instrument configurations
(e.g., HV values,  threshold values and trigger schemes) but also on background levels and  solar activity.

The housekeeping data contains instrument information such as working modes,
values of currents and voltages of the low voltage power supplies, MAPMT high voltage levels, 
module temperature readouts, values of the signal discriminating thresholds, module counting rates as well as status of the CT and counters and timestamps of the injected and executed commands.
The platform data contains locations, speed and orientation  of the TG2 space-lab updated every 500 milliseconds.
These data are not generated by the POLAR instrument but provided by the TG-2 space-lab itself.

\begin{table*}[htp]
\centering
\caption{Raw data received at the POLAR data  center at PSI. }
\label{tb:types}
 \resizebox{\textwidth}{!}{%
\begin{tabular}{cp{6cm}p{2cm}cp{2cm}}
\toprule
Data Type         & Contents     & Daily average size           & Source     &Number of files                      \\
\hline
Science      & Trigger pattern, module trigger bits, ADC channels, etc.    & 50 GB  & OBOX    & 5     \\
Housekeeping & Temperature values, voltage and currents, module count rates etc. &  20 MB  & OBOX and IBOX &5 \\
Platform & Location, speed and orientation of the space-lab, etc.    &   60 MB  & TG2 Space-lab&30 \\
\bottomrule
\end{tabular}
}
\end{table*}

The POLAR science data and housekeeping data are collected and temporarily stored in the Payload
Data Handling Unit (PDHU) on-board the space-lab.
They are sent back to ground  stations via
S-band  or through relay satellites within a few hours.
Data received by different ground stations is collected and processed by the
TG-2 space-lab Payload Operation and Application Centre (POAC).
After the processing, 
POLAR data is binary data. It has the same structure
as the data transmitted to the PDHU.
The data are written into different files according to their types.
The naming of the files indicate their levels and data types.

The raw POLAR data is then transmitted to a storage server
at the Institute of High Energy Physics (IHEP) via  FTP protocol.
A server at the PSI data center periodically synchronizes with the raw server at IHEP by using
Linux rsync.
The data arrive at PPDC with a typical delay of 12 hours following
its receipt at POAC.

When a new file is received, its MD5 checksum  is compared
with the one stored in a file generated by the IHEP data server.
If they are not same, it means that the file
is not identical to the one in the IHEP server.
A synchronization of the file  will start again.
After this consistency check, the filename,
the size, the directory, the data acquisition start time and the stop time,
data arrival time, the MD5 checksum, the data type and  some other information for each raw data file is inserted into a database table called ``raw-data''.

\section{Database}

\begin{table*}[htp]
\centering
\caption{Example of columns in the selected database tables. }
\label{tb:dbtables}
 \resizebox{\textwidth}{!}
 {%
\begin{tabular}{p{0.1\linewidth}p{0.2\linewidth}p{0.7\linewidth}}
\toprule
\#         & Table name    & Columns                      \\ \hline\
1 & module-configuration & HV, threshold, data compression mode,
command creation time,
command injection time, command execution time, etc. \\ \hline
2 & raw-data &  filename, directory, start time, stop time, arrival time, MD5, filesize, etc.\\ \hline
3 & level-0 & input filenames, run number, creation time, output file directory, start time, stop time, etc. \\
\hline
4 & GRBs & burst name, start time, stop time, T90, RA, DEC, peak flux,  total counts,
theta, phi, latitude, longitude, quick-look directory, etc. \\
\bottomrule
\end{tabular}
}
\end{table*}

POLAR will receive more than 40,000 raw data files over three years of operation.
Every raw dataset is followed by several higher level data files and hundreds of plots
generated by the quick-look software.
In order to manage such a huge number of data files and
also simplify information sharing between different programs and between different system branches,
a relational MySQL database   was developed  at PPDC.
It consists of 39 tables and archieves about 500 columns in total.
Each table contains different data collections of the processed information.
Table \ref{tb:dbtables} shows examples of columns for four database tables: ``module-configurations'', ``raw-data'', ``level-0'' and ``GRBs''.
The table ``module-configurations'' records  configurations for each POLAR module.
Its columns contain among other the high voltage values, threshold values and module IDs.
They allow a reconstruction of the detector module configuration at any given moment.
The  table ``raw-data'' records information of received raw data files such as
filenames,  file directories, data acquisition start and stop time, files arrival time and their sizes.
The  database table ``level-0''  contains columns that maintain information on the level-0 data
such as run numbers, input filenames, data start time and stop time, and output file directories.
The table ``GRBs'' stores the start time, duration, location, incident angle, spacecraft location, peak flux, 
number of detected events and the Quick-Look outputs for each detected GRB.
The database is heavily used by both the data processing pipeline and the web applications. 
It is used to store virtually all information that must to be exchanged between different system
branches or software packages.
A vast majority of the columns in the database tables are filled and updated by the data processing pipelines autonomously
i.e. without any human manual intervention.

\section{Data flow and data levels at PPDC}
\begin{figure*}[!htb]
\begin{center}
\includegraphics[width=0.98\textwidth]{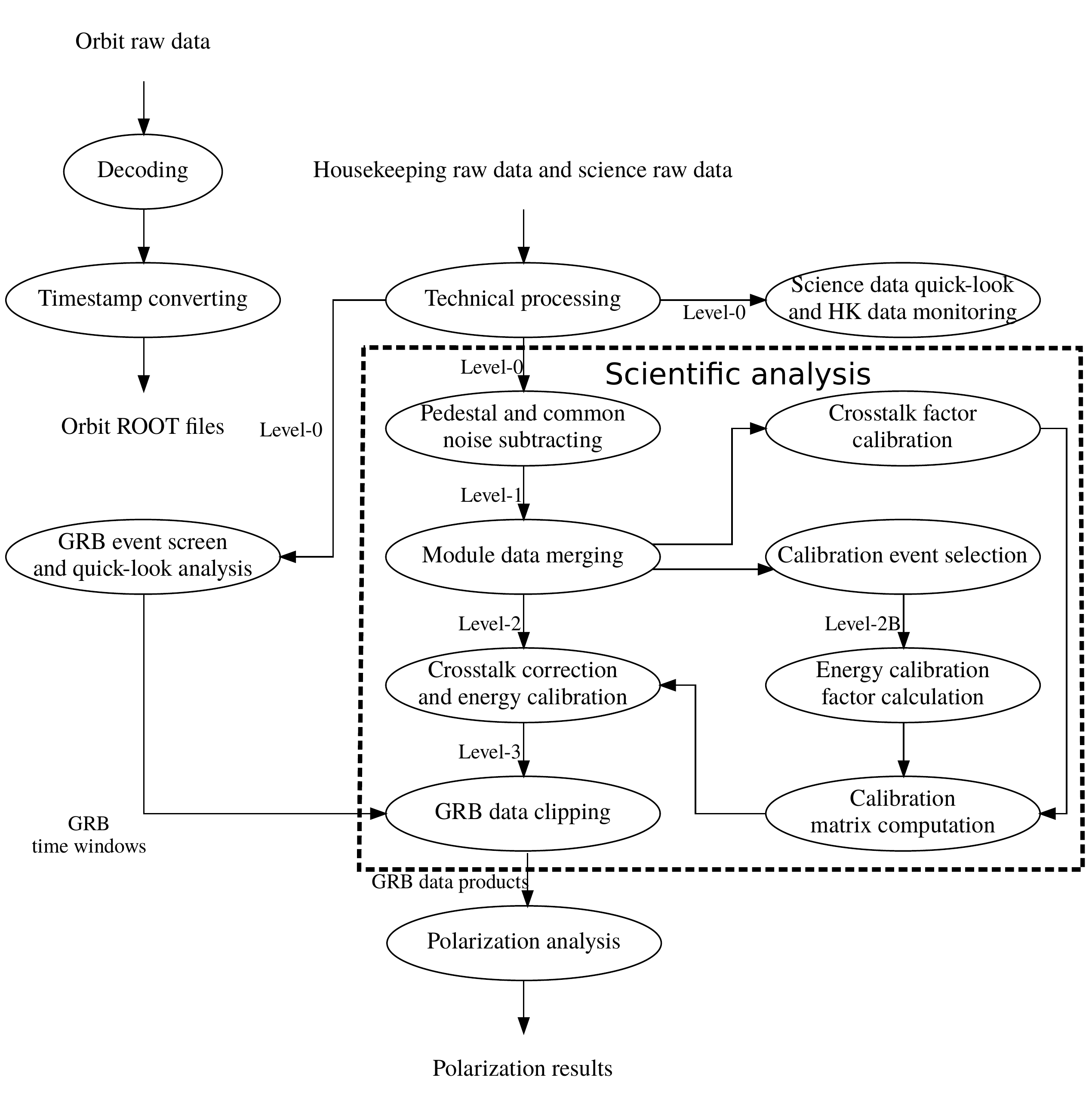}
\caption{Data flow and data levels at the PSI POLAR data center. }
\label{fig:dataflow}
\end{center}
\end{figure*}


Three different types of raw data arriving at the PPDC are processed in 
two different pipelines as shown in
Fig.~\ref{fig:dataflow}.
The housekeeping and science data are processed in the same pipeline, whereas data from the platform uses a different pipeline.
Platform data files are initially decoded to obtain data physical values that are written to the ROOT files.

Raw data from the science and housekeeping packets are processed separately in another pipeline.
It starts with the technical processing of the data. In the next step the raw data files are converted into the
level-0 data. They are further processed by the software responsible for  housekeeping monitoring 
and by the quick-look analysis application. The last one generates a thorough view of the instrument status.
Further processing aims to create the level-1 datasets. 
As the first step all pedestals and common mode noise values are subtracted for each physical event.
Subsequently all hits belonging to the same physical event but recorded at different modules are merged.
Additionally, the merged events are attributed by all relevant housekeeping data
such as module temperatures and HV values.  
They are written to the new data files making the level-2 datasets.
Afterward, data merging by the coincidence algorithm described in Ref.~\cite{hlastro2018}
is applied to the  level-2 datasets. This step also enables extraction of the in-flight calibration events.
They are written to the new data files assigned as level-2B.
Finally, level-3 datasets contain data after crosstalk corrections and application of energy calibration factors.
Events selected from the level-3 datasets for a  particular GRB form the initial GRB data products.
They are used for further analysis of the polarization.
More details on the data processing pipelines are introduced in the next sections.

\section{Platform data processing}

When a new data file from the platform is received at PPDC, it triggers an autonomous start of the 
dedicated program that decodes and interprets its contents. The binary data contain 
spacecraft location, attitude and speed as well as the GPS timestamps with increments every half second.  GPS timestamps are converted into Unix-timestamps in which the leap seconds are also taken into account. After processing, the platform data is written to the ROOT format files.
The reason for adopting the ROOT format is due to its clever compression capability and high computing efficiency. The data start and stop time, data processing time, input filename and ID of the output file of each processing
are recorded in a dedicated database table.


\section{Pipeline for housekeeping and science data processing}

\subsection{Technical processing}
\label{sec:tech}
Before  the data can be used for detailed scientific analysis it requires certain technical processing. The corresponding processing program is autonomously triggered to start when 
a new science data file  as well as its corresponding housekeeping data exists on the server. 
Technical processing uses
information recorded in the  ``raw-data'' database table.
The processing itself involves a series of steps including checking of CRC code, data  decoding and interpretation,
conversion of GPS timestamps to Unix-timestamps, synchronization of module timestamps, and calculations of count rates.
It also attaches the housekeeping data  (e.g. temperatures, threshold values and high voltage values)
to each recorded event.
The science data of 25 modules, the trigger data, the housekeeping data, the pedestal events 
and module packet alignment information of each processing run are written to different ROOT files.
A common run number is assigned to all output files from the same processing run.
The file naming convention of the output files is shown in Table \ref{tb:techprocess}.
\textless RUN\textgreater \  is a 5-digit number representing the run number and  \textless MID\textgreater \  ranging from 0 to 24 represents the ID of the detector module.
The data produced by technical processing runs are called the level-0 data.

For the housekeeping packets the level-0 data contains the same amount of information as the raw data.
The level-0 module science data  contains event-by-event science data as well as the corresponding 
housekeeping information: PMT high voltage level, threshold values and temperature sensor readouts stored for each event.
As noted above, this information is extracted from the housekeeping packets.
Attachment of the housekeeping data to the events greatly simplifies event selections for conditions 
such as selected ranges of temperatures, PMT high voltages or thresholds.

The trigger rate files contain counting rates from 25 modules and total counting rates calculated for every 250 ms.
They are widely used by the web-based light curve viewers.

\begin{table*}[htp]
\centering
\caption{Level-0 data types and the file naming convention.
}
\label{tb:techprocess}

\begin{tabular}{ccp{7cm}}
\toprule
\#         & filename      &Content                 \\ \hline\
1 & sci\_\textless MID\textgreater\_\textless RUN\textgreater.root   & Module science data and auxiliary data from housekeeping  \\
2 & hkct\_\textless RUN\textgreater.root  &  Housekeeping data, updated every two seconds \\
3 & trig\_\textless RUN\textgreater.root &  Trigger data\\
4 & sci\_rate\_\textless RUN\textgreater.root  & trigger rates for the whole instrument and 25 different modules, updated every 250 ms\\
5 & ped\_\textless RUN\textgreater.root  &Pedestal events\\
6 & extra\_\textless RUN\textgreater.root  & Module packet alignment information \\
\bottomrule
\end{tabular}

\end{table*}

It is worth mentioning that the FITS format which is a standard data format in astronomy is not used here.
This is due to its low efficiency in treatment of data written on an the event-by-event basis and involvement of many obsolete structures. 
After each processing run the processing log, input filenames, run number, output file directory and the 
data start and stop times are inserted into the database table ``level-0''.
The level-0 data is normally generated within 30 minutes after corresponding raw data packets are received.

\subsection{GRB signal identification and quick analysis software}
\begin{figure*}[htb]
\begin{center}
\includegraphics[width=0.8\textwidth]{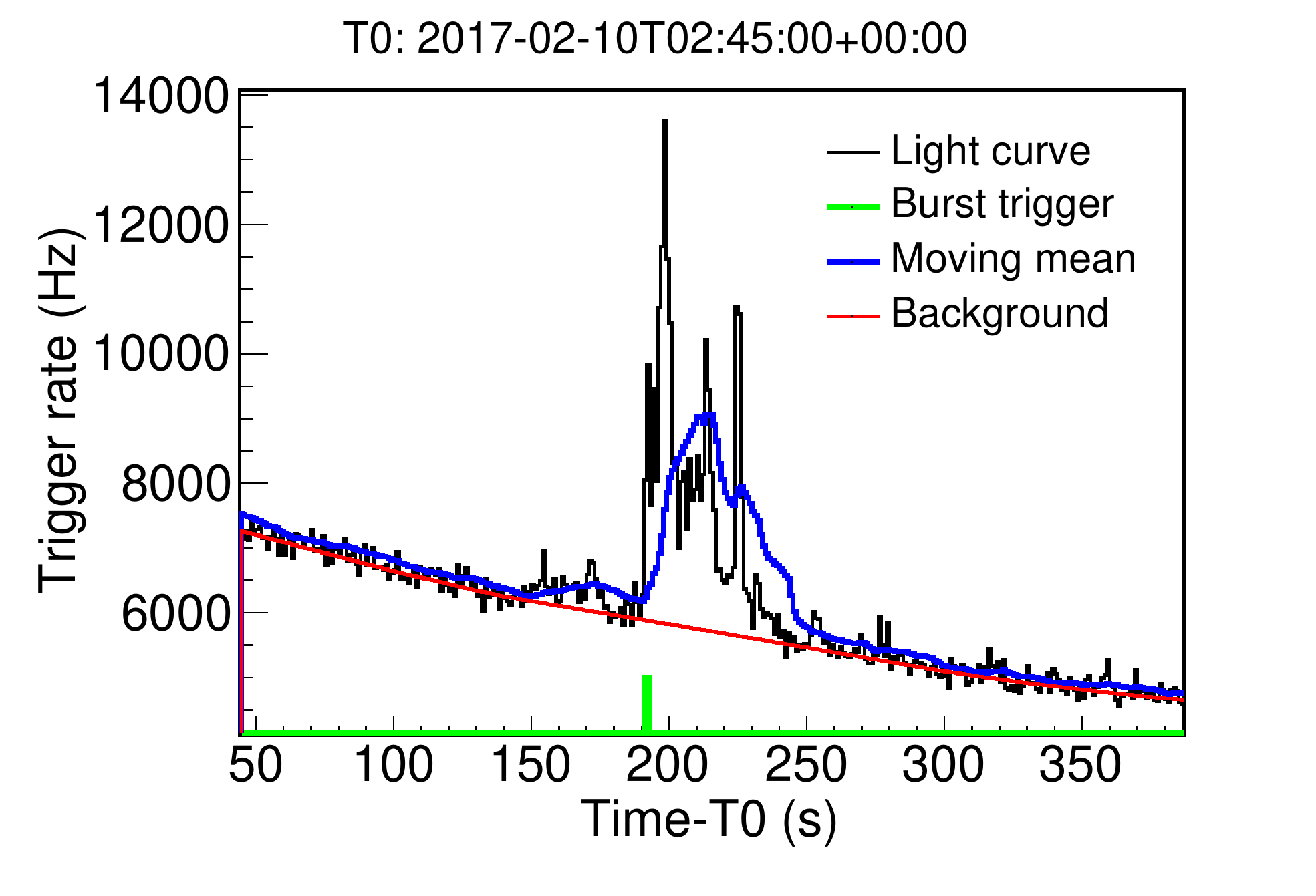}
\caption{Light curve measured for the GRB 170210A together with the moving average and background curves.}
\label{fig:lc}
\end{center}
\end{figure*}

GRB signals are characterized by sudden increases of accumulated counts (i.e., peaks) observed in the light curves such as 
e.g. count rate time series.
As POLAR hardware and onboard software do not identify such signals they must be identified in an offline analysis.
A simple algorithm  based on the moving average is adopted for this purpose.
The algorithm achieves this by computation of the moving mean counting rate $\bar{x_i}$ and 
its moving standard deviation $\sigma_i$. It is done for every data point with a moving window of $n_w$  data points.
A burst is considered to be detected at the moment at which the count rate $x_i$  is above the
threshold $T$ defined by  $ T=\bar{x_i}+ a \cdot\sigma_i$, where $a$ is a parameter.

The method relies on two parameters, $n_w$ and $a$. 
They were optimized and tested with the data from confirmed GRBs.
In most cases we chose $n_w=20$ and $a=3.5$. 
Fig. \ref{fig:lc} shows the light curve measured for 
the GRB 170210A together with its moving average.
The GRB is identified at the time point equal to
191 s when the threshold of 3.5 $\sigma$ is observed above the moving mean value. 
The running average used a window with the width of 20 data points.
The background was calculated with the Sensitive Nonlinear Iterative Peak (SNIP) 
clipping algorithm implemented in the ROOT package \cite{bkgalg}. 
The method is very robust because of its dynamic threshold knowledge learned 
from the previous data points.

If a burst is detected, the level-0 data in the GRB time window is clipped. 
The clipped data is further processed in a number of steps:
\begin{enumerate}[1)]
\item light curve preparation using the clipped data.
\item background estimation based  on the Sensitive Nonlinear Iterative Peak (SNIP) algorithm from ROOT (see Fig.~\ref{fig:lc}).
\item light curve background subtraction.
\item calculation and storage of GRB parameters: start and stop times, duration (T90), total number of events, peak flux.
\item calculation of platform data (TG2 position and orientation) and burst incident angles.
\item processing of the clipped data with the housekeeping data monitor and the science data quick-look software (see the next section).
\item automatic publishing of processed information and plots on several internal dynamic web pages 
	\footnote{ 
		\url{http://polar.psi.ch/pdc/burstqklook.php}}\footnote{\url{http://polar.psi.ch/pub}}.
\item generating an email notice containing the above information and links to created web pages; sending it to all collaborators.
\item continuous listening to the notices from the Gamma-ray Coordinates Network (GCN), which
distributes the GRB and transient locations determined by several satellites\cite{gcn}.
Storage of received external GRB triggers in a dedicated database.
Comparison of POLAR triggers with external triggers recorded in the database.
If any external trigger matches with the POLAR  one and the GRB is in the POLAR's field of view the burst is confirmed
and published on the database public web pages.
\item  Preliminary calculation of the GRB modulation curves with estimations of the polarization for the GRB 
detected either automatically or implemented manually. The results are published on the GRB data quick analysis web page.
\end{enumerate}

It is important to note that  the database experiences heavy use  sharing of the data between
numerous different routines 
such as external triggers extracted from the GCN notices,
burst start and stop times, data file run number, 
clipped data directories, parameters calculated for
GRBs and quick analysis output directories.
The whole management of the quick-look plots and quick 
analysis results with their web pages also relies on this database.

\subsection{Housekeeping data monitoring and science data quick-look}

Both the housekeeping data monitor and the science data quick-look use
the level-0 housekeeping data as their inputs.
The housekeeping monitoring software detects if the recorder value of any housekeeping parameter
(e.g., voltage,  current, count rate, temperature, etc.) 
 violates its limits which are stored in a dedicated database table.
This allows for comprehensive determination of the POLAR health status.
An email notice is sent to all registered POLAR collaborators if any failure or malfunctioning
of the instrument is detected. 
An operation request is then submitted to the TG-2 operation 
center by the instrument operation board when it is necessary. 

Plots of all housekeeping variables are also created and written to both PNG image
files and ROOT files by the above software.
Important instrument status and the indexing information of the above outputs are recorded in the database.
They are used later by the scientific analysis software as well as web applications.
The science quick-look analysis program  may produce thousands of
different plots for each run. For example it generates maps of hit patterns and noisy channels,
counting rates, pedestal values, hit non-uniformity maps as
well as physical event energy spectra and much more.
All these data are necessary to properly assess the performance and the status of the detector.
The processing time, data run number, instrument health status and output indexing information
of each  processing run are recorded in several dedicated database tables.

The plots are created in the PNG format and grouped run by run.
They can be retrievable with dynamically generated web pages according to users'
query conditions \footnote{
	\url{http://polar.psi.ch/pdc/qklook.php}}\footnote{ 
\url{http://polar.psi.ch/pdc/hkmon.php}}.
All processing steps listed above are usually completed within 2 hours
after reception of the raw data.
A daily report containing e.g. detector health status, 
data quantity, daily light curves as well as the web page URLs
is sent to POLAR collaborators via E-mail.

\subsection{Standard scientific analysis}
The standard scientific analysis includes a number of generally defined
steps as shown in Fig.~\ref{fig:dataflow}. In the first step it corrects for detector effects
and converts the level-0 data to the next level.
During analysis the accumulated pedestal events are used for pedestal position calculation \cite{silvio},
while for physical events, both the pedestals and common noise subtractions are performed.
The data with subtracted pedestals and common noise values is defined as the level-1 data.
It has the same structure as the level-0 data.
In the next step all hits belonging to the same physical event (within the electronic resolution) and recorded by
different modules are merged. For this purpose one uses the module packet alignment
information from the level-0 auxiliary data.
Additionally, the merged events contain all relevant housekeeping
data, e.g., such as module temperatures and HV values, etc.
The merged events are written to  new data files making the level-2 dataset.

It should be noted that POLAR does not have a separate calibration mode. 
All data required for its calibration are acquired in the normal operation mode.
The in-flight calibrations use four weak radiation sources sending annihilation photons in opposite directions (collinear). Such events can be treated as  background data. Real calibration data taking is conducted using proper HV values. After the merging, collinear annihilation photons are selected if the HV values are suitable for in-flight calibration. The events selected with a proper algorithm that, e.g., takes into account positions of the sources, are considered as being created by the internal positron sources. They are written to new data files assigned as level-2B.
Compared to the level-2 data, all events in the level-2B dataset contain extra information used for extraction 
of the energy calibration factors.
The extracted energy calibration factors are written to the ROOT files. 
Final energy calibration factors are calculated by extrapolation based on the gain dependence on the HV values.
In the next step, some background events are selected for the crosstalk factor calibration.
The crosstalk correction method is similar to the one described in Refs.~\cite{hlastro2016} and \cite{silvio}. 
When both of the crosstalk matrices and energy calibration factors  are ready, the full detector response 
matrices are calculated and also stored in the ROOT files.
They are used to apply  crosstalk corrections and energy calibrations for events in the level-2 dataset.
After the above processing, the level-2 dataset is converted to the level-3 data.

Another set of routines is assembled to prepare the pre-selected data necessary for polarization analysis.
They include  the POLAR instrument configurations, calibration data,
and different levels of the data clipped for both signals and background.
Details of the polarization analysis of GRBs detected by POLAR will be published in future papers.

All relevant information about each of the above processing
runs (and respective GRBs candidates), i.e., input filenames, run numbers, processing time,
output filenames, reports and processing status etc. is recorded in the database.
With its help each processing run knows if its
inputs are ready and properly placed in predefined locations.
More details about the steps and methods applied for the standard scientific analysis processing are described in Refs.~\cite{hlastro2018,silvio, hlastro2016}.

\section{Web applications}

\begin{figure}[htb]
\begin{center}
\includegraphics[width=1.0\textwidth]{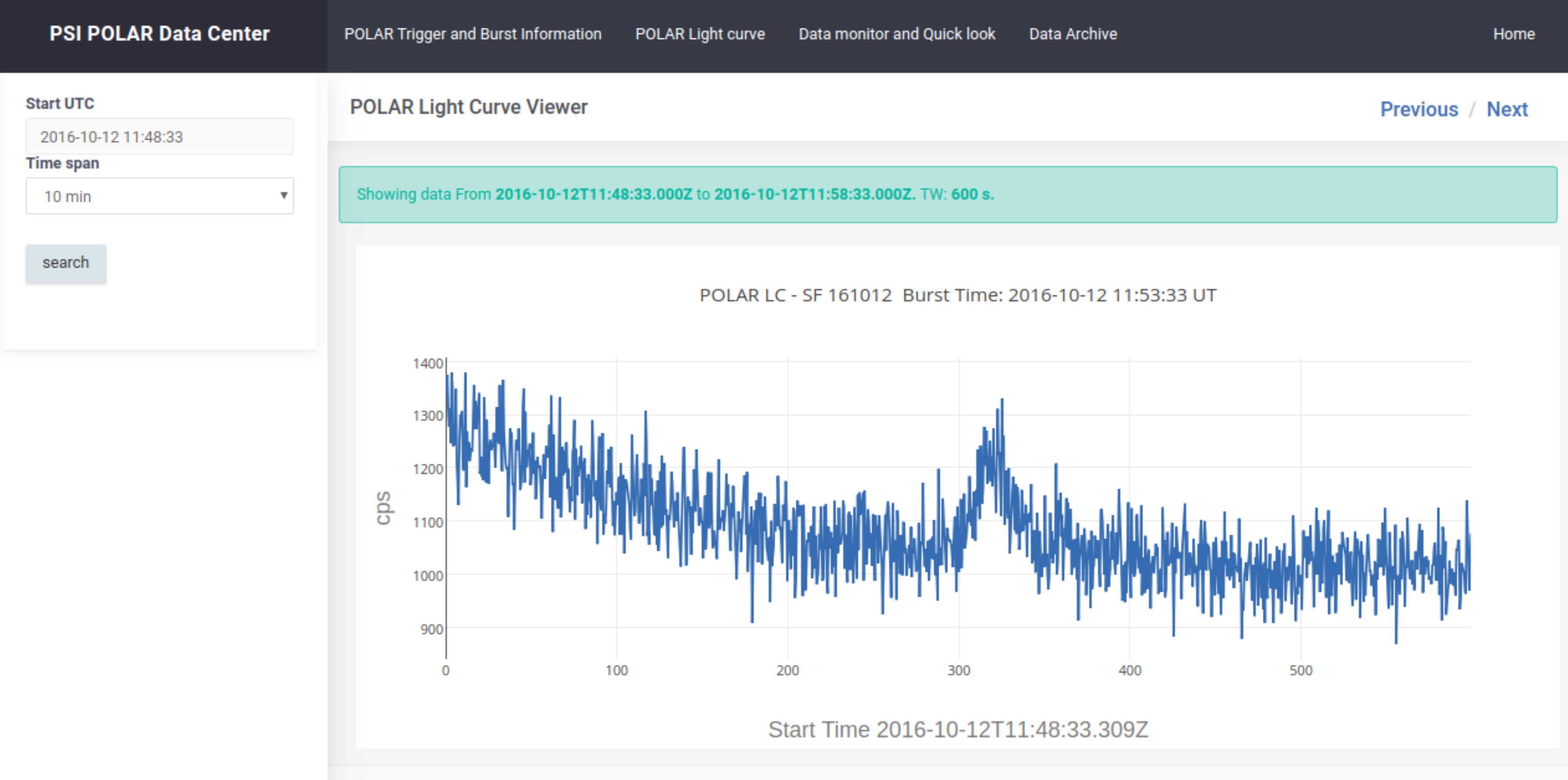}
\caption{Web-based light curve viewer developed at PPDC.
	It is accessible via \url{http://polar.psi.ch/pub/lc.php}.}
\label{fig:weblcviewer}
\end{center}
\end{figure}

\begin{figure}[htb]
\begin{center}
\includegraphics[width=1.0\textwidth]{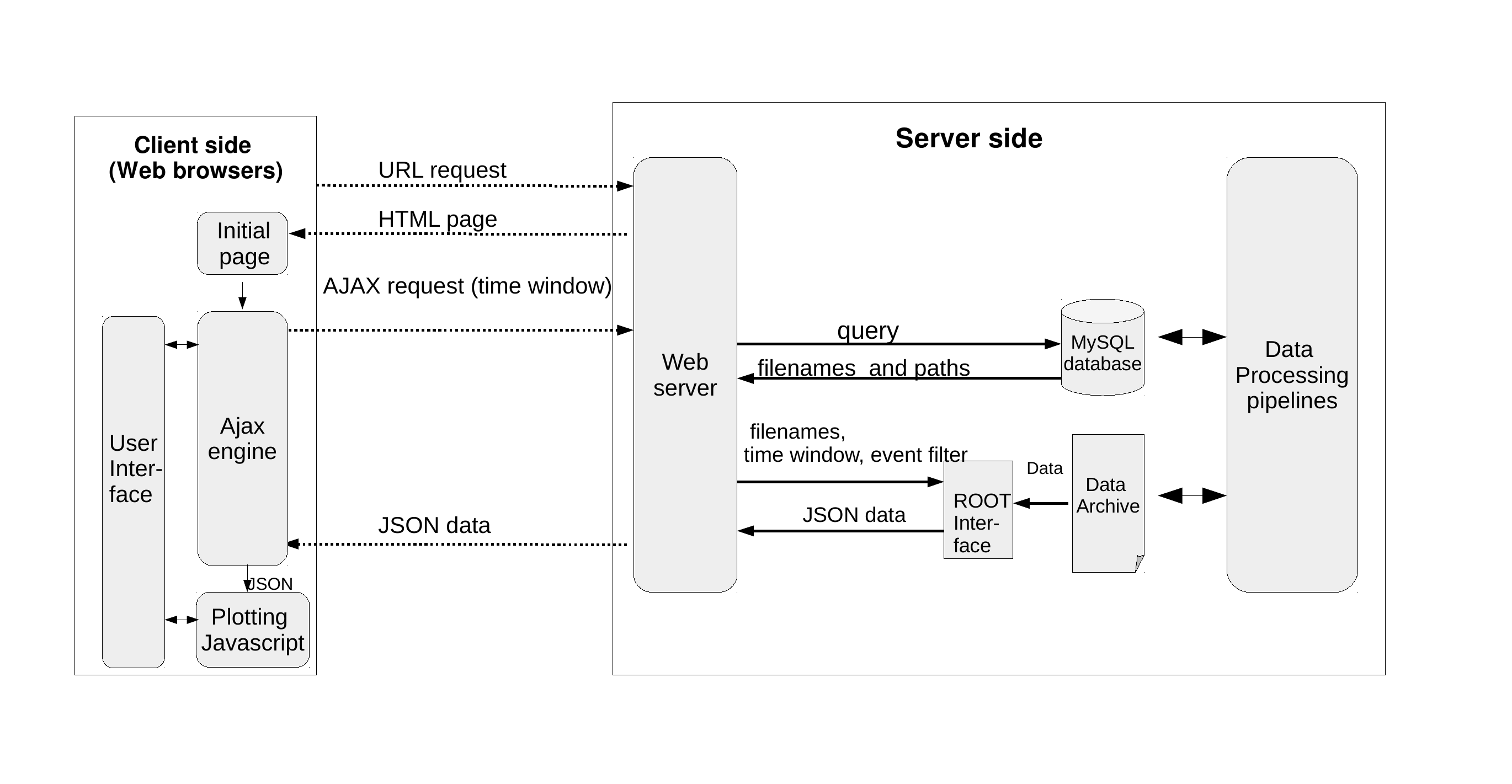}
\caption{
Client-side and server-side mechanisms of a light curve viewer developed at the POLAR data center.
}
\label{fig:web}
\end{center}
\end{figure}
Web-based applications are selected for the graphical interfaces at the data center.
This allows for the advantages of clear cross-platform usability and wide access for
internet browsers.
As an example, Fig.~\ref{fig:weblcviewer} shows the web-based light curve viewer mentioned in
Section \ref{sec:tech}.  
The client side and the server side interact and operate as shown in Fig.~\ref{fig:web}.
The details are given below:
\begin{enumerate}[a)]
\item User interface (UI) is prepared when the page is opened by the browser.
\item The user sends a request form with  time window information from the UI.
\item The Javascript captures the user command and a data request is sent to the server by the AJAX engine.
\item The web server interacts with the database using PHP.
\item If the data exists on the server, data indexing information (data filenames, their paths etc.)
is sent to the web server.
\item The web server requests data from the ROOT interface.
\item The data is retrieved by the ROOT interfacing routine and converted to JSON streams.
They contain all the data required to plot the light curve.
\item The JSON streams are sent to the client-side web browser.
\item The AJAX engine receives the streams and the light curve is plotted with Javascript on the browser.
\end{enumerate}

More than 50 different web applications were purposely developed
for the needs of the POLAR data center web UI.
Their mechanisms are similar to those described above.
The web applications provide graphical interfaces to retrieve and view the POLAR light curves, housekeeping monitoring
and science quick analysis plots, GRB data quick-look results, space-lab orbit data, etc. 
Moreover, they provide interfaces to access various data products stored
on the server including proper management of the database tables.
The web applications are accessible through the link \href{http://polar.psi.ch}{http://polar.psi.ch}.

\section{Summary}
POLAR is a space-borne hard X-ray Compton polarimeter dedicated to precise measurements of the polarization
of hard X-rays emitted by transient sources such as Gamma Ray Bursts and Solar flares 
in the 50 keV to 500 keV energy range.
It was launched into space on September 15th, 2016 on-board the Chinese Space Laboratory TG-2 for up to a 3-year long observation period.
A dedicated POLAR data center was established at PSI  in order to store and process the huge amounts of data
from its operation in space.  
The database consists of data files, software packages and web applications including user interfaces. 
The data files contain both raw and preprocessed data files up to the level-3. 
The center was designed to work in fully autonomous mode with only minimal intervention from human operators. 
The concept proved to be successful and it has been running in a continuous manner for more than
two years without human intervention.
The center offers a wide range of applications allowing for science and housekeeping data monitoring and preprocessing 
as well as a powerful user interface with advanced quick-look software tools and extensive summary tables. 
To date POLAR has detected more than 50 GRBs and slightly more solar flares.
Most of the detected GRBs were identified  automatically by the offline GRB identification software
developed at the center. Polarization studies of the detected GRBs are ongoing.

\bibliographystyle{model1-num-names}

\bibliography{reference}

\end{document}